# *Where there's a will there's a way:* ChatGPT is used more for science in countries where it is prohibited

Honglin Bao[1*], Mengyi Sun[2*], Misha Teplitskiy[3†]

[1] Data Science Institute, University of Chicago
[2] Simons Center for Quantitative Biology, Cold Spring Harbor Laboratory
[3] School of Information, University of Michigan

\* These authors contributed equally
† To whom correspondence should be addressed: Misha Teplitskiy (tepl@umich.edu)

**Abstract.** Regulating AI is a key societal challenge, but which regulation methods are effective is unclear. This study measures the effectiveness of restricting AI services geographically, focusing on ChatGPT. OpenAI restricts ChatGPT access in several countries, including China and Russia. If restrictions are effective, ChatGPT use should be minimal in these countries. We measured use with a classifier based on distinctive word usage found in early versions of ChatGPT, e.g. "delve." We trained the classifier on pre- and post-ChatGPT "polished" abstracts and found it outperformed GPTZero and ZeroGPT on validation sets, including papers with self-reported AI use. Applying the classifier to preprints from Arxiv, BioRxiv, and MedRxiv showed ChatGPT was used in about 12.6% of preprints by August 2023, with 7.7% higher usage in restricted countries. The gap appeared before China's first major legal LLM became widely available. To test the possibility that, due to high demand, use in restricted countries would have been even higher without restrictions, we compared Asian countries with high expected demand (where English is not an official language) and found that use was higher in those with restrictions. ChatGPT use was correlated with higher views and downloads, but not citations or journal placement. Overall, restricting ChatGPT geographically has proven ineffective in science and possibly other domains, likely due to widespread workarounds.

# Introduction

The emergence of generative AI has elicited calls for its regulation (Bengio et al., 2024; Biden, 2023). Policymakers may seek to regulate the speed of generative AI development and adoption, align it better with societal needs, and reduce disruptions to labor markets, national security, and the information environment (Cohen et al., 2024; Weidinger et al., 2022; Whittaker et al., 2018). Policymakers in generative AI-producing countries may also seek to prevent the technology from being used by their global competitors (Alper, 2024; Lee, 2018), while those in AI-consuming countries may want to limit diffusion to cultivate domestic development and achieve political goals (Ray, 2023). These concerns also extend to the scientific community, where many fear generative AI can harm academic integrity (Blau et al., 2024). As discussions over whether and how to regulate generative AI gain steam, it is important to identify the policy instruments that are and are not effective. At present, empirical studies of policy effectiveness are missing.

Here, we contribute such evidence, measuring the effectiveness of geographical restrictions on generative AI using the case of ChatGPT, in scientific research. OpenAI, the company behind the world's most prominent large language model (LLM) ChatGPT[1], prohibits access to it in several countries including China, Russia, and Iran (see Table S1 in the Appendix for the full list). To create accounts for early versions of ChatGPT users were required to have IP addresses and phone numbers in a permitted country. On July 9 2024, OpenAI implemented further actions to expand on its existing policy of blocking users in unsupported territories, in response to the continuous detection of API traffic originating from China[2]. In addition to this "supply-side" restriction, some countries, notably China, have additional "demand-side" restrictions (Ray, 2023). Work on digital piracy suggests that restrictions and bans are unlikely to eliminate it, but may raise the cost of use enough to discourage at least the casual if not professional user (Adermon & Liang, 2014; Reimers, 2016). Accordingly, if the geographical restrictions are effective, ChatGPT use should be lower in countries where it is prohibited. Is this the case?

Measuring the use of technology like ChatGPT, with or without legal access, is challenging. Some prior work has used surveys (Bonney et al., 2024), with one study finding that 31% of postdocs reported using LLMs (Nordling, 2023) and a staggering 76% of academics reported using some form of generative AI (MacGregor, 2024). However, surveys may suffer from reporting biases since in many contexts generative AI tools are not ethically or legally approved (*More than Half of Generative AI Adopters Use Unapproved Tools at Work*, 2023). Others have used off-the-shelf classifiers like GPTZero and ZeroGPT (Cheng et al., 2024; Latona et al., 2024; Picazo-Sanchez & Ortiz-Martin, 2024), but these detectors have not been convincingly validated on scientific texts, and as we show below, their accuracy is quite low. Here, we draw on findings that early LLM-generated scientific texts overrepresent certain "fingerprint" words, such as "delve" and "intricate" (Gray, 2024; Kobak et al., 2024; Liang, Izzo, et al., 2024), and construct a classifier that uses such features to identify ChatGPT use in abstracts (See Materials and Methods). Using a sample of Arxiv, BioRxiv, and MedRxiv abstracts, we use ChatGPT to "polish" them and train a classifier on the before vs. after versions. We validate the classifier on held-out abstracts from these preprint servers and abstracts from papers with and without AI use declarations. We then apply the classifier to scientific preprints from the aforementioned sources and track use over time and by country. The domain of science, and preprints in particular, are attractive for measuring ChatGPT use because scientists generally make their outputs public to get recognition and other rewards and, unlike published papers, preprints become public relatively quickly. Lastly, academic science is an increasingly global endeavor, with a large portion being produced in countries without legal ChatGPT access, such as China.

---

[1] ChatGPT refers to the LLM GPT-3.5 Turbo. The application programming interface (API) for accessing this model, as well as the chatbot built on it, are not available to the countries listed in Table S1.

[2] https://www.moneycontrol.com/news/world/openai-to-cut-access-to-tools-for-developers-in-china-and-other-regions-12755873.html and https://www.theregister.com/2024/06/25/openai_unsupported_countries/

Although our analysis focuses on ChatGPT (GPT 3.5) specifically, some LLMs like GPT 2.0 were available before its launch, and many more, like Google Gemini, have become available subsequently. Our research design does not rule out the use of these alternatives, but two factors make this unlikely to affect our conclusion. First, in the study period, ChatGPT was by far the most popular LLM (see "Clean and full periods" below) and, second, the alternatives generally had the same geographical restrictions[3] as ChatGPT.

# Materials and Methods

## Corpora

**BioRxiv and MedRxiv** (biology, medicine, and related research). We downloaded metadata of preprints posted between March 2019 and March 2024 using the BioRxiv and MedRxiv APIs. The metadata included the DOI, abstract, submission date, subfield, and, if published, journal. Additionally, we used the OpenAlex API to obtain author affiliation and country.

---

[3] For example, Google Gemini is prohibited in China, Iran, and Russia (*Where You Can Use the Gemini Web App - Gemini Apps Help*, 2024). Gemini's supporting country list is also available at: https://support.google.com/gemini/answer/13575153?hl=en

This resulted in a dataset containing 233,997 preprints and 932,209 unique authors across 76 biomedical-science subtopics, such as "bioinformatics." For conciseness, we refer to BioRxiv and MedRxiv together as simply "BioRxiv."

**Arxiv** (computational science and related research). We used Arxiv metadata from Kaggle (https://www.kaggle.com/datasets/Cornell-University/arxiv), which includes the DOI, field category, abstract, publishing journal, and posting date of each preprint. To clearly distinguish the posting date relative to the introduction of ChatGPT, we included preprints with only one version. We randomly sampled approximately 2,000 preprints per month from April 2019 to March 2024, resulting in a dataset of 106,357 preprints from 319,383 unique authors across 8 fields including computer science, mathematics, physics, electronic engineering, quantitative biology, quantitative finance, statistics, and economics. Using the Arxiv API, we downloaded PDF files of all preprints in this dataset and used Grobid[4], a widely used package for processing scientific PDFs, to extract the country and author affiliations from each preprint[5]. If the country information was not extractable from the PDF files, we queried OpenAlex to obtain the country code based on author affiliations.

**Elsevier**: In August 2023 Elsevier announced a policy requiring explicit declarations in papers that used generative AI tools for writing and suggested specific declaration text (*The Use of AI and AI-Assisted Technologies in Writing for Elsevier*, n.d.). We identified 70 recent papers published by around 50 journals that acknowledged AI writing assistance using the suggested text. For each journal in the set, we identified up to 125 "non-adopter" papers published from 2017 to 2021[6].

## Citation, journal, and attention data

For both Arxiv and BioRxiv preprints, we collected the yearly citation data and journal impact factor (if published) using OpenAlex API. 48.86% of BioRxiv preprints and 28.03% of arXiv preprints were published. In addition, for BioRxiv only we collected attention statistics (number of abstract views, PDF downloads, and full paper views online within the first six months of the preprint's posting) by scraping the preprint website.

To make these outcomes more comparable across time and subfield, and to account for skew (see Appendix 5, Figure S5), we transformed each outcome into percentiles, calculated within the relevant time period and subfield (self-reported by authors upon submission) and divided the percentile by 100, so that 1.0 represents the highest rank. For attention, the time period used for normalization was posting year, for journal impact factor publication year, and for citations posting month.

---

[4] https://grobid.readthedocs.io/en/latest/. Accessed 2024-06-13.
[5] We found that for this dataset OpenAlex had relatively poor coverage of authorship data, and used it only as a secondary option.
[6] Some journals had fewer than 125 papers matching the inclusion criteria.

## Clean and full periods

We distinguish between two time periods based on the level of certainty that the LLM used was ChatPGT and not its alternatives. OpenAI officially announced the introduction of ChatGPT on November 30, 2022. The resulting popularity of ChatGPT appeared to be unexpected even by those inside OpenAI (Roose, 2023). In response, Google first released Google Bard (later renamed Gemini) on February 06, 2023[7] and in the subsequent months non-ChatGPT LLM tools like Google Gemini (Bard), LLaMA, and Claude AI became available. We refer to the period from November 30, 2022, to February 5, 2023, as the "clean period" because ChatGPT 3.5 was the only major public-facing LLM available. We refer to the clean *and* the subsequent period as the "full period." Although non-ChatGPT LLMs were in principle available in the latter part of the full period, ChatGPT remained the predominant LLM: Google searches for it accounted for approximately 90% of the total search volume for LLMs from June 2023 to June 2024 (see Figure S1 in the appendix). It is also important to take into account LLMs developed around the world, such as Baidu's Ernie Bot, developed for the Chinese context and released on March 16, 2023. These models were initially available to a strictly limited group of users through invitation codes only and released to the general Chinese public after August 2023 (Yang, 2023). Overall, there is near-certainty that the LLM used, if any, in the clean period was ChatGPT, and that certainty declines slowly over the full period.

## Measuring legal access

We measure legal access with OpenAI's official list of supported countries. This list is dynamic. For instance, ChatGPT was banned in Italy from March 31, 2023 to April 28, 2023. We constructed a time-varying access indicator for each country (see Table S1 in the Appendix). We exclude countries with fewer than 10 preprints before and after ChatGPT in our corpora and no legal access from the analysis on access and ChatGPT use.

## Classifier

To identify ChatGPT use we draw on recent work showing that ChatGPT-modified text overrepresents certain words like "delve" and "intricate," *i.e.* fingerprint words (Gray, 2024; Kobak et al., 2024; Liang, Zhang, et al., 2024). We focus on identifying ChatGPT 3.5 as this is the LLM behind "ChatGPT," which brought LLMs into the mainstream and supercharged the "A.I. arms race" (Roose, 2023).

---

[7] https://blog.google/technology/ai/bard-google-ai-search-updates/

We train a machine learning classifier on preprint abstracts to predict whether an abstract was ChatGPT-assisted. We use the full text of abstracts (without the newline character '\n') for training and adjust tokens' weights using the TF-IDF (Term Frequency-Inverse Document Frequency) vectorizer, which captures the importance of unique tokens across the corpus. We do not limit the tokens used to only fingerprint words like "delve" because other linguistic features may help distinguish ChatGPT-assisted from unassisted writing. For example, prior work found that ChatGPT texts use more determiners, conjunctions, and auxiliary relations, while human texts use more exclamation marks, question marks, and ellipses (Tang et al., 2024).

The classifier is an ensemble of four common text classifiers: Support Vector Machine (SVM with a linear kernel) and Logistic Regression, and the other two suitable for nonlinear cases: Random Forest (with 100 estimators) and Gradient Boosting Decision Tree (with 100 estimators). Different classifiers have different logics: SVMs focus on maximizing boundaries; logistic regression focuses on the linear relationship between features and labels; random forests rely on the importance ranking of features (*e.g.*, fingerprint words); and gradient boosting decision trees iteratively adjust the error. Each classifier returns a probability of the text label 1. We average these four probabilities to generate the final label: if the average probability is greater than 0.5, the text is labeled 1, and otherwise 0.

We validated our classifier in two ways. First, we tested it on held-out parts of Arxiv/BioRxiv preprints before and after ChatGPT "polishing." The second scenario uses the Elsevier declaration dataset. Our simple classifier outperforms commercial tools like ZeroGPT (https://www.zerogpt.com/) and GPTZero (https://gptzero.me/), which show the lowest bias against non-native English writing samples in the context of TOEFL essays (Liang et al., 2023) among all existing tools, in both scenarios.

**Training and testing sets:** For the Arxiv/BioRxiv training set, we have a set of prompts that mimic how people use ChatGPT to polish their writing (see Appendix 6). We collected 5,000 samples from BioRxiv before the release of ChatGPT on 2022-11-30 (so no papers would have used ChatGPT), as label 0. Each abstract then was paired with a prompt randomly to generate a ChatGPT-polished version as label 1 for the training set of BioRxiv corpus. We constructed the Arxiv training corpus including 20,000 label 0 and 20,000 label 1. This corpus was larger than BioRxiv because we found that the token distribution in the Arxiv corpus is more difficult for the classifier to learn. Similarly, we collected 2000 label 0 (1000 from countries with access and 1000 from those without access) and 2000 label 1 for

Arxiv/BioRxiv as the overall testing sets but we sampled from the overall testing sets to randomize the proportion of label 1 to test classifiers for both corpora from countries with vs. without access (see below). For the Elsevier corpus, we ran abstracts from those journals from 2017 to 2021 (label 0) through ChatGPT with a randomly selected prompt from Appendix Section 6 to obtain the polished version (label 1) as the training set. We use those 70 abstracts with explicit AI usage declarations as label 1 and collect an additional random 70 abstracts from the same journals from 2017 to 2021, as label 0. We use these three testing sets to evaluate the domain-trained classifier's effectiveness in detecting text modified by ChatGPT. We have ensured that testing sets do not overlap with training sets.

**Measuring performance**: We are interested in determining how accurate the predicted proportion is and how the classifier performs across different proportions of true (*i.e.* predetermined) ChatGPT usage. We drew 10,000 random numbers evenly distributed from 0 to 1 to set the true proportion of ChatGPT use in that iteration. For each proportion $i$, we sampled $k$ label 1 and $k \times (\frac{1-i}{i})$ label 0 instances from the testing set to ensure the proportion of label 1 in the sampled set meets the set proportion, where $k$ is a random number in [0, the total number of label 1's]. If the number of required label 0 instances was larger than the total label 0's in the testing corpus, we selected all label 0 and recalculated the actual proportion of label 1. We conducted these 10,000 simulations separately for corpora from countries with access and those without access, to check whether errors were larger in one group or the other. For each iteration we calculated standard classifier evaluation metrics including accuracy, F1 score, recall, precision, and ROC-AUC (see Results and Appendix 7). We also calculated the predicted proportion based on the sampled set and the gap (absolute value) between the predicted and true proportions.

# Results

## Fingerprint words

First, we identify fingerprint words that ChatGPT overrepresented by calculating the change in the fraction of abstracts including each word in Arxiv and BioRxiv abstracts and their ChatGPT-polished versions from the training dataset. Figure 1 displays the month dynamics in the full corpus of four words with the highest increase (panel A) and decrease (panel B) in usage, with the first vertical dashed lines indicating the introduction of ChatGPT and the second indicating the introduction of Google Bard and other LLMs subsequently. The increased words corroborate prior work (Gray, 2024; Liang, Zhang, et al., 2024), while the decreases indicate that it is important to include in fingerprint words those that ChatGPT underrepresents. While we do not identify the features the ensemble classifier relies on *directly*, we check whether they are plausibly these usage changes. Specifically, we measured the usage of these eight words in texts the classifier predicted as "1" and "0" and confirmed they are over- and underrepresented in those labels as expected.

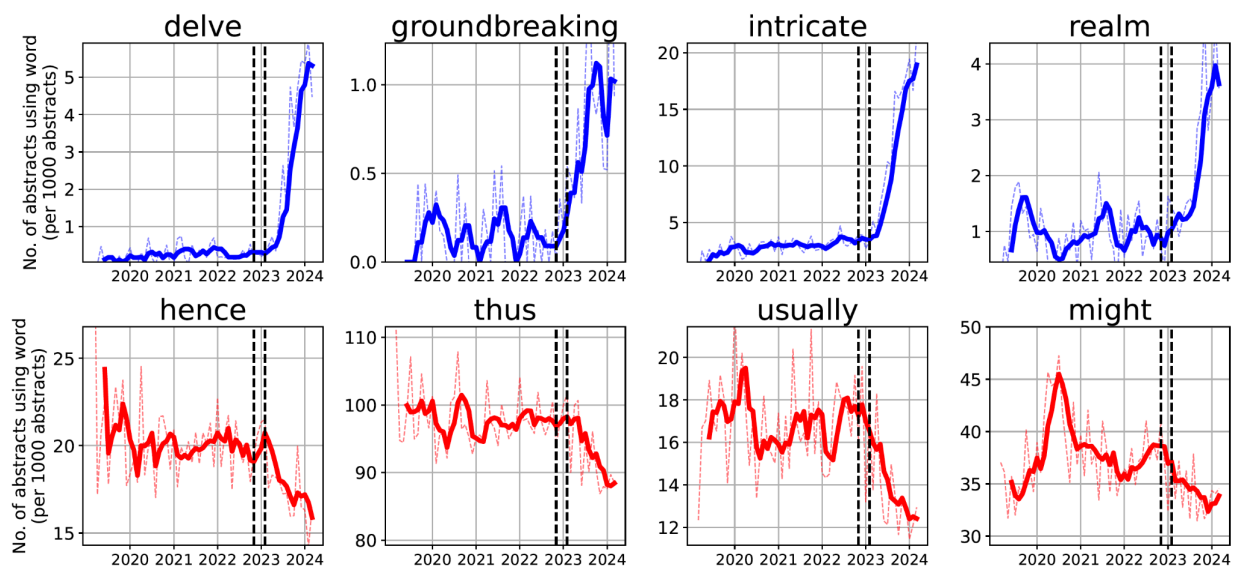

**Figure 1**. *Four words with the largest increases (top row) and decreases (bottom row) in usage comparing the original vs. ChatGPT polished abstracts in the training set. The first vertical dashed line indicates the introduction of ChatGPT and the second one other LLMs. The blue/red dashed lines indicate the monthly proportion of abstracts that include the word, while the blue/red solid lines show the corresponding four-month rolling averages.*

## Comparing our classifier with off-the-shelf LLM detectors

For each of our three corpora – Arxiv, BioRxiv, and Elsevier – we constructed 10,000 test sets where the proportion of ChatGPT-modified text was randomized between 0.0 and 1.0 (see Data and Methods) and calculated the accuracy of our ensemble classifier and two widely-used LLM detectors, ZeroGPT and GPTZero. The results are displayed in Figure 2. Our ensemble classifier outperformed both off-the-shelf detectors on all three corpora. Furthermore, the error rates for our classifier were not significantly higher in countries without legal access, none of which are predominantly English-speaking (see Figure S6 and Tables S1 and S2 in the appendix), alleviating concerns about biases against non-native English writers found in off-the-shelf detectors (Liang et al., 2023). While off-the-shelf detectors seldom misclassify human-written text as ChatGPT-generated (low false positives), they frequently fail to identify a significant amount of ChatGPT-modified text (high false negatives) as a result of the lack of domain-specific training. This suggests that there is no one-size-fits-all method for detecting LLM-modified text, and even a simple classifier trained on a domain-specific corpus can achieve higher domain-specific accuracy.

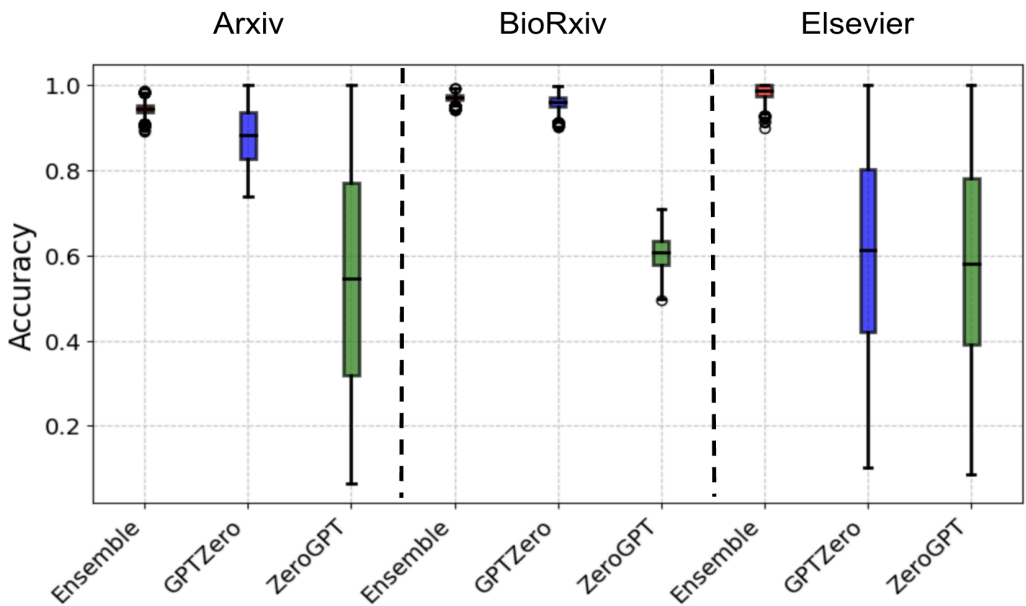

*Figure 2.* Comparison of the accuracy of the ensemble classifier and two popular LLM detectors in three corpora. The ensemble method we developed outperformed GPTZero and ZeroGPT in accurately identifying ChatGPT-modified text in Arxiv (left), BioRxiv (middle), and Elsevier corpora (right).

## Legal access and ChatGPT use

We apply the classifier to the Arxiv and BioRxiv corpora to track use over time. The predictions are illustrated in Figure 3. The dashed lines indicate the monthly usage and the solid lines four-month rolling averages. The figure indicates that the use of ChatGPT in the *pre*-ChatGPT period was above zero, seemingly a contradiction. These estimates most likely reflect the classifier's false positive rate, although we cannot rule out some use of pre-ChatGPT LLMs, such as GPT2, that overrepresent similar tokens as ChatGPT. Crucially, the likely erroneous pre-ChatGPT estimates are similar for countries without legal access (2.35% on BioRxiv and 2.77% on Arxiv) and with access (2.01% on BioRxiv and 4.45% on Arxiv). The small asymmetry in errors in Arxiv makes our subsequent comparison of ChatGPT use in the post-ChatGPT period more conservative. In general, Panel A shows that preprints from countries *without* legal access show *higher* use of ChatGPT in the clean and full periods, with the gap growing over time. In both the ArXiv and BioRxiv corpora over 80% of preprints from countries without legal access are from China (see Appendix section 3 for the distribution). Therefore, we divide the non-access group into "China" and "not China" and show estimated usage for each corpus in Panels B and C. In general, use is highest in preprints from China. It is important to note that this high use appeared before Aug 31 2023, when Ernie Bot, the first major LLM legal in China, was released to the public.

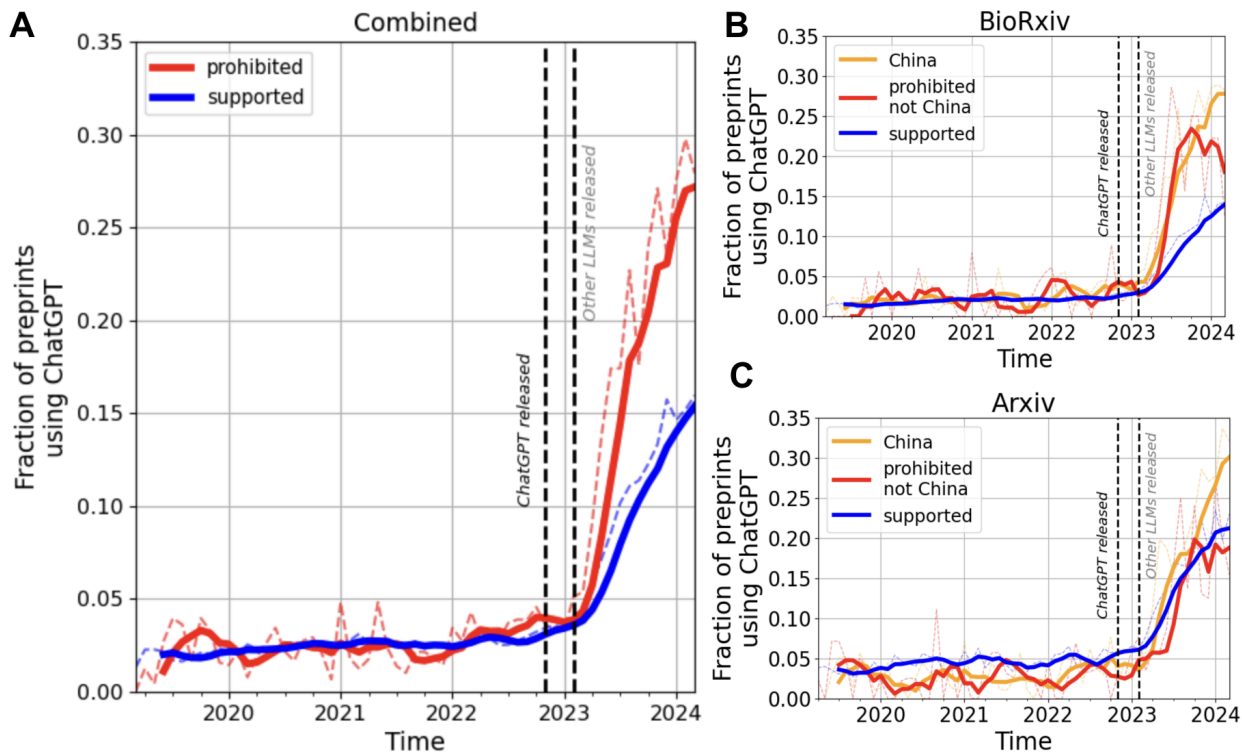

**Figure 3.** *Predicted use by countries with and without access. (a) Overall usage trend of ChatGPT on BioRxiv and Arxiv. Countries without legal access (red) have a higher usage rate of ChatGPT compared to countries with legal access (blue). (b) Usage trend of ChatGPT on BioRxiv. China (orange) has the highest usage, followed by other countries without access (red), and countries with access (blue). (c) Usage trend of ChatGPT on Arxiv. China (orange) has the highest adoption, followed by other countries without access (red), and countries with access (blue).*

We test whether these visual differences are statistically significant using OLS regressions that control for whether the preprint is in a computational subfield (for BioRxiv corpus, the subfields are bioengineering, bioinformatics, genetic and genomic medicine, genomics, health informatics, systems biology) or its subject category (for Arxiv corpus), specified as below:

$$ChatGPT\_use_{i,t} = \beta_0 + \beta_1 \times without\_access_{i,t} + \beta_2 \times subject_i + \beta_3 \times months\_post\_ChatGPT_t + \epsilon_{i,t}$$

where $ChatGPT\_use_{i,t}$ is the prediction (1 or 0) of paper $i$ using ChatGPT after its release, $\beta_0$ is the intercept, $without\_access_{i,t}$ denotes whether *none* of the authors had legal access to ChatGPT, $months\_post\_ChatGPT_t$ is a set of year-month indicators for each post-ChatGPT month, and $\epsilon_{i,t}$ is the error term. The key parameter of interest is $\beta_1$, which reflects the change in predicted ChatGPT use if the authors are from *without_access* countries. We estimate these regressions using data from the clean and full periods separately. We cluster the standard errors within the last author as the majority of subjects/topics in the Arxiv and BioRxiv corpora follow the norm that the last author is the senior author. Our findings remain robust when operationalizing "without access" using the locations of the first and last authors (see section 9 in the Appendix).

As shown in Table 1A, for BioRxiv, the estimated $\beta_1$ is 0.000 (SE=0.009, p=0.961) in the clean period and +0.079 (SE=0.007, p<0.001) in the longer full period. Table 1B shows that for Arxiv, the estimated $\beta_1$ is -0.021 (SE=0.013, p=0.091) in the clean period and +0.022 (SE=0.008, p<0.01) in the full period. Overall, the more precise full period estimates show higher use from countries without access. The less precise clean period estimates are mixed, showing lower use in Arxiv and higher use in BioRxiv, but neither is statistically significant. Overall, the results challenge the expectation that restrictions substantially reduce use: ChatGPT use in countries without legal access was not significantly lower than those with access in the clean period and became substantially higher afterward.

*Table 1A. OLS models predicting* use as a function of legal access and other features in the BioRxiv/MedRxiv corpus. Left column shows predictions for the full period and right column for the clean period. Standard errors are clustered at the last author level.

|  | ChatGPT use (BioRxiv) | |
| --- | --- | --- |
|  | Full | Clean |
| without access | 0.079*** | 0.000 |
|  | (0.007) | (0.009) |
| intercept | 0.014*** | 0.026*** |
|  | (0.003) | (0.003) |
| is computational | 0.058*** | 0.013* |
|  | (0.004) | (0.005) |
| year-month indicators | Y | Y |
| Observations | 62847 | 8359 |
| $R^2$ | 0.033 | 0.001 |
| Adjusted $R^2$ | 0.032 | 0.000 |
| Residual Std. Error | 0.281 (df=62828) | 0.166 (df=8354) |
| F Statistic | 104.098*** (df=18; 62828) | 1.521 (df=4; 8354) |
| Note: | *p<0.05; **p<0.01; ***p<0.001 | |

**Table 1B. OLS models predicting** *use as a function of legal access and other features in the Arxiv corpus. Left column shows predictions for the full period and right column for the clean period. Standard errors are clustered at the last author level.*

|  | ChatGPT use (Arxiv) | |
| --- | --- | --- |
|  | Full | Clean |
| without access | 0.022** | -0.021 |
|  | (0.008) | (0.013) |
| intercept | 0.053*** | 0.055*** |
|  | (0.009) | (0.011) |
| computer science | 0.075*** | 0.029** |
|  | (0.006) | (0.010) |
| economics | 0.031 | 0.020 |
|  | (0.043) | (0.085) |
| electronic engineering | 0.004 | 0.007 |
|  | (0.012) | (0.023) |
| mathematics | -0.085*** | -0.016 |
|  | (0.007) | (0.014) |
| physics | -0.075*** | -0.022 |
|  | (0.006) | (0.012) |
| quantitative biology | 0.042 | -0.043* |
|  | (0.022) | (0.020) |
| quantitative finance | 0.112** | -0.000 |
|  | (0.043) | (0.082) |
| statistics | 0.010 | -0.012 |
|  | (0.014) | (0.021) |
| year-month indicators | Y | Y |
| Observations | 15299 | 2220 |
| $R^2$ | 0.066 | 0.010 |
| Adjusted $R^2$ | 0.064 | 0.005 |
| Residual Std. Error | 0.347 (df=15273) | 0.234 (df=2208) |
| F Statistic | 40.917*** (df=25; 15273) | 1.943* (df=11; 2208) |

*Note:* *p<0.05; **p<0.01; ***p<0.001

## ChatGPT use among countries with similar expected demand

Countries in the comparisons above may differ not only on legal access, but demand for ChatGPT. Accordingly, those results do not rule out the possibility that restrictions *did* reduce use in prohibited countries from a higher (counterfactual) no-restriction level. To test this possibility we compare countries where we expect demand for ChatGPT to be similar — Asian countries where English is not an official language, *i.e.* those that can benefit substantially from ChatGPT's writing capabilities. Regression analyses indicate that among these countries, use was not associated with legal access in the clean period ($\beta_1$= -0.007, SE=0.011, p=0.515), and was again higher among those without access in the full period ($\beta_1$= +0.022, SE=0.008, p<0.01) (see Appendix 11). Overall, we do not find evidence supporting the possibility that use in restricted countries would have been substantially higher without restrictions.

## ChatGPT use and academic outcomes

To illuminate the incentives for ChatGPT use, we measured its associations with five academic outcomes. For the BioRxiv and Arxiv corpora, the outcomes are citations (to the preprint *and* published version, if published) and the impact factor of the publishing journal. For the BioRxiv corpus only we also considered attention measures of preprints, including the number of six-month abstract views, PDF downloads, and full paper online views, as attention data is unavailable in Arxiv. All measures are normalized as percentile values within time periods and subfields to account for their skewed nature, as described in the Materials and Methods section. We then estimated panel models with last author fixed effects, which account for the academic outcomes last authors typically experience, due to their skills or resources for example. Given that baseline, the models measure how the outcomes vary with ChatGPT use and whether the preprint was posted before or after ChatGPT introduction, and are specified as follows:

$$attention/impact\_factor/citations_{i,t} = \beta_1 \times is\_post\_ChatGPT_{i,t} + \beta_2 \times ChatGPT\_use_{i,t} \times is\_post\_ChatGPT_{i,t} + \alpha_i + \epsilon_{i,t}$$

$\alpha_i$ represents the last author fixed effect and $\epsilon_{i,t}$ the error term. We estimate this model using data only from authors who were last authors on at least one preprint posted before *and* after the introduction of ChatGPT. The key parameter of interest is $\beta_2$, which reflects the change in outcomes for preprints that used ChatGPT after its release. Our results are robust when using first author fixed effects (see Appendix 10).

As shown in Table 2, for the attention change in BioRxiv corpus, the estimated $\beta_2$ is +0.020 (SE=0.009, p=0.023) for full paper online view, +0.022 (SE=0.008, p<0.01) for PDF downloads, +0.024 (SE=0.009, p<0.01) for abstract views in the longer full period. The estimated $\beta_2$ is -0.007 (SE=0.029, p=0.800) for full paper online view, -0.007 (SE=0.030, p=0.813) for PDF downloads, -0.012 (SE=0.032, p=0.716) for abstract views in the clean period. Regarding the change in citations and journal placements in the Arxiv and BioRxiv corpora, neither the clean nor full periods show statistically significant effects of ChatGPT use. Collectively, these findings indicate that while the use of ChatGPT increases readers' attention to the papers, it does not affect their citations or the quality of their final publication.

Table 2A. Panel OLS regressions predicting attention (full paper online view, PDF downloads, and abstract view) accumulated in six months after posting as a function of predicted ChatGPT use, time, and last author fixed effects. The first three columns use the full period and the last three the clean period. Standard errors are clustered at the last author level. Attention measures are normalized by converting to percentile ranking within papers from the same field and time.

|  | Full View Full Period | PDF Full Period | Abstract Full Period | Full View Clean Period | PDF Clean Period | Abstract Clean Period |
|---|---|---|---|---|---|---|
| used ChatGPT× post-ChatGPT | 0.020* (0.009) | 0.022** (0.008) | 0.024** (0.009) | -0.007 (0.029) | -0.007 (0.030) | -0.012 (0.032) |
| post-ChatGPT | -0.009*** (0.002) | -0.007** (0.002) | -0.006* (0.002) | -0.040*** (0.005) | -0.052*** (0.005) | -0.053*** (0.005) |
| Last author-FE | Y | Y | Y | Y | Y | Y |
| Observations | 70916 | 71401 | 71392 | 20264 | 20424 | 20424 |
| N. of groups | 15700 | 15735 | 15734 | 4355 | 4363 | 4363 |
| $R^2$ | 0.000 | 0.000 | 0.000 | 0.005 | 0.009 | 0.008 |
| Residual Std. Error | 0.005 (df=55214) | 0.004 (df=55664) | 0.004 (df=55656) | 0.018 (df=15907) | 0.024 (df=16059) | 0.024 (df=16059) |
| F Statistic | 9.349*** (df=15702;55214) | 7.095*** (df=15737;55664) | 5.345** (df=15736;55656) | 39.415*** (df=4357;15907) | 69.242*** (df=4365;16059) | 62.404*** (df=4365;16059) |

Note: *p<0.05; **p<0.01; ***p<0.001

Table 2B. *Panel OLS regressions predicting* citations and journal placement (if published in a journal) as a function of predicted ChatGPT use, time, and last author fixed effects. The first four columns use the full period and the last four the clean period. Standard errors are clustered at the last author level. Both citations and impact factors (IF) are normalized by converting to percentile ranking within papers from the same field and time.

|  | IF Full Period BioRxiv | Citation Full Period BioRxiv | IF Full Period Arxiv | Citation Full Period Arxiv | IF Clean Period BioRxiv | Citation Clean Period BioRxiv | IF Clean Period Arxiv | Citation Clean Period Arxiv |
|---|---|---|---|---|---|---|---|---|
| used ChatGPT× post-ChatGPT | 0.000 (0.017) | -0.008 (0.005) | 0.006 (0.041) | 0.011 (0.007) | -0.044 (0.051) | -0.012 (0.027) | -0.100 (0.088) | -0.015 (0.038) |
| post-ChatGPT | -0.014*** (0.004) | -0.032*** (0.002) | -0.021 (0.011) | -0.023*** (0.004) | -0.008 (0.008) | -0.024*** (0.005) | -0.022 (0.022) | -0.021* (0.009) |
| Last author-FE | Y | Y | Y | Y | Y | Y | Y | Y |
| Observations | 20458 | 98721 | 1803 | 18705 | 6241 | 20430 | 543 | 3413 |
| N. of groups | 5087 | 21500 | 656 | 5683 | 1532 | 4363 | 211 | 1022 |
| $R^2$ | 0.001 | 0.005 | 0.003 | 0.003 | 0.000 | 0.002 | 0.007 | 0.002 |
| Residual Std. Error | 0.007 | 0.017 | 0.012 | 0.012 | 0.005 | 0.011 | 0.019 | 0.011 |
| F Statistic | 6.096** df=5089; 15369 | 177.698*** df=21502; 77219 | 1.766 df=658; 1145 | 19.567*** df=5685; 13020 | 1.101 df=1534; 4707 | 15.202*** df=4365; 16065 | 1.224 df=213; 330 | 2.617 df=1024; 2389 |

Note: *p<0.05; **p<0.01; ***p<0.001

# Discussion

This paper investigates whether geographical restrictions on the use of ChatGPT were associated with lower actual use. Focusing on the domain of scientific research and using ChatGPT for writing assistance, we developed a classifier for identifying ChatGPT-modified abstracts and applied it to a large set of preprints across time and geography. The results show striking and quickly-growing adoption of ChatGPT by the scientific community, reaching about 12.6% of abstracts by August 2023. Relative to countries where ChatGPT is permitted, its use was generally *higher* in countries where it is *prohibited*. For example, by August 2023 estimated use in preprints from China was about 22.29% and legal-access countries 11.07%. This difference appeared before legal alternatives like Ernie Bot became widely available in China.

Observed use is a function of not only restrictions but also demand. Accordingly, we considered the possibility that demand in restricted countries like China exceeds that from other countries, so use there would have been even higher without restrictions. We tested this possibility by comparing countries with similar expected demand – Asian countries like China and South Korea where English is not an official language – but that differed on restrictions. Contrary to the possibility, use was higher among restricted countries in this subset.

We also explored how ChatGPT use was associated with important academic outcomes, like citations and attention. In the BioRxiv corpus, ChatGPT-using preprints showed about two percentiles higher views of abstract, PDF, and online full text, while associations with citations and journal impact factor for all corpora were less precisely estimated and not statistically significant. One potential explanation for the inconsistency across outcomes is that, unlike views, citations and placement in top journals are more zero-sum. Overall, the data suggest an association between use and attention, but more research is needed to establish precision and causality.

Developing and validating our classifier also provided the opportunity to test the performance of popular off-the-shelf LLM detectors in the scientific domain. Figure 2 shows that both ZeroGPT and GPTZero perform poorly relative to our relatively simple but task-specific classifier. As the use of such detectors for studying scientific texts increases, their performance limitations should be kept in mind.

The results should be interpreted with several limitations in mind. First and foremost, we do not measure ChatGPT use directly, for example via administrative data. Our measure infers use from token frequencies in the abstract. This approach has many merits but prevents us from ruling out two alternative explanations of the main finding, even if they are unlikely. First, the training data for the classifier was generated by applying ChatGPT with particular prompts, such as "Please polish the following abstract. Be clear" (Appendix 6). It is possible that the prompts we chose were unrepresentative or geographically biased. Future empirical research on prompts is needed to assess the validity of the prompts we deployed. Second, we focused on language in the abstract because it is a plausible proxy for using ChatGPT for other tasks and because of high data availability. However, if researchers in countries with legal access, relative to those without it, use ChatGPT for various tasks but *not* abstract writing, then our results may underestimate use in the former and perhaps even obscure the real effects of restrictions. Empirical research on how exactly researchers use LLMs is also much needed. Third, we cannot rule out the use of alternative LLMs, particularly after ChatGPT's introduction. As noted in the Introduction and Materials and Methods, their use in the several months after November 2022 should not materially affect our conclusions because of their relatively low popularity and geographical restrictions that mirrored OpenAI's. Lastly, the study is of only one domain – science. How the results generalize to other domains is unclear. Scientists may be more savvy or motivated to find workarounds for restrictions than workers in other industries, perhaps due to ChatGPT's capabilities in translation and the importance of publishing research in English ("Scientific Publishing Has a Language Problem," 2023). Measuring the effects of restrictions across industries is another important avenue for future work.

# Conclusion

As the question of *whether* to regulate generative AI is hotly debated, the question of *how* to regulate it is perhaps just as pressing. One key empirical case can already be evaluated – the attempt to restrict the use of ChatGPT, the world's leading LLM, to a set of permitted countries. Using scientific preprints from computing, engineering, and biomedicine, we find no evidence that restrictions reduce use. On the contrary, use of ChatGPT for scientific writing was generally higher in countries where it was prohibited, and we find no evidence for the possibility that use there would have been even higher without restrictions.

Interestingly, ChatGPT use was associated with only modest benefits to attention, and no statistically significant benefits to citations and journal placement. One possible interpretation is that the alleged benefits of ChatGPT for science are overstated. Another, and perhaps more plausible, interpretation is that its effects help scientists increase the amount of outputs, but not their quality.

In the end, the results beg the question of why the seemingly strong restrictions have proven ineffective. While we lack direct measures, publicly available information reveals that after the launch of ChatGPT, an ecosystem arose quickly to provide workarounds for individuals without legal access. In restricted countries approved accounts can be bought and sold online, appropriately located IP addresses and phone numbers can be bought as well, and entrepreneurs and enthusiasts offer intermediary proxy servers (Zhao, 2023). The apparent swiftness and effectiveness of these workarounds services are important in considering how the intent of AI regulations may be undermined by non-compliance.

# Appendix to

# *Where there's a will there's a way: ChatGPT is used more for science in countries where it is prohibited*

Honglin Bao, Mengyi Sun, Misha Teplitskiy

**Table of Contents**

# 1. Clean vs. full period

Google released Google Bard on February 6, 2023 and soon renamed it as Gemini. To our knowledge this was the first significant public-facing competitor to ChatGPT. Google Trends data indicates that starting from the week of February 5, 2023, there was an increase in searches for large language models (LLMs) other than ChatGPT (Figure S1, panel a). Despite this, ChatGPT still dominates search traffic, accounting for over 90% of all searches of major LLMs from June-18 2023 (the start of trackable Google search data for 'ChatGPT" as software) to June-15 2024 (panel b).

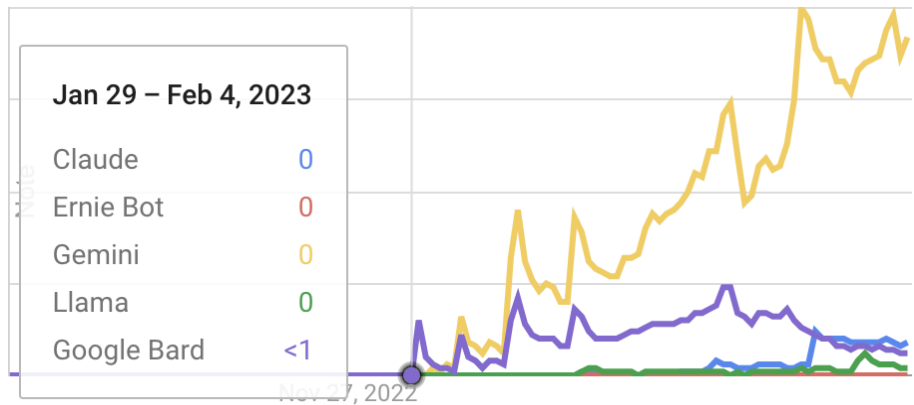

(a) The end of the clean period.

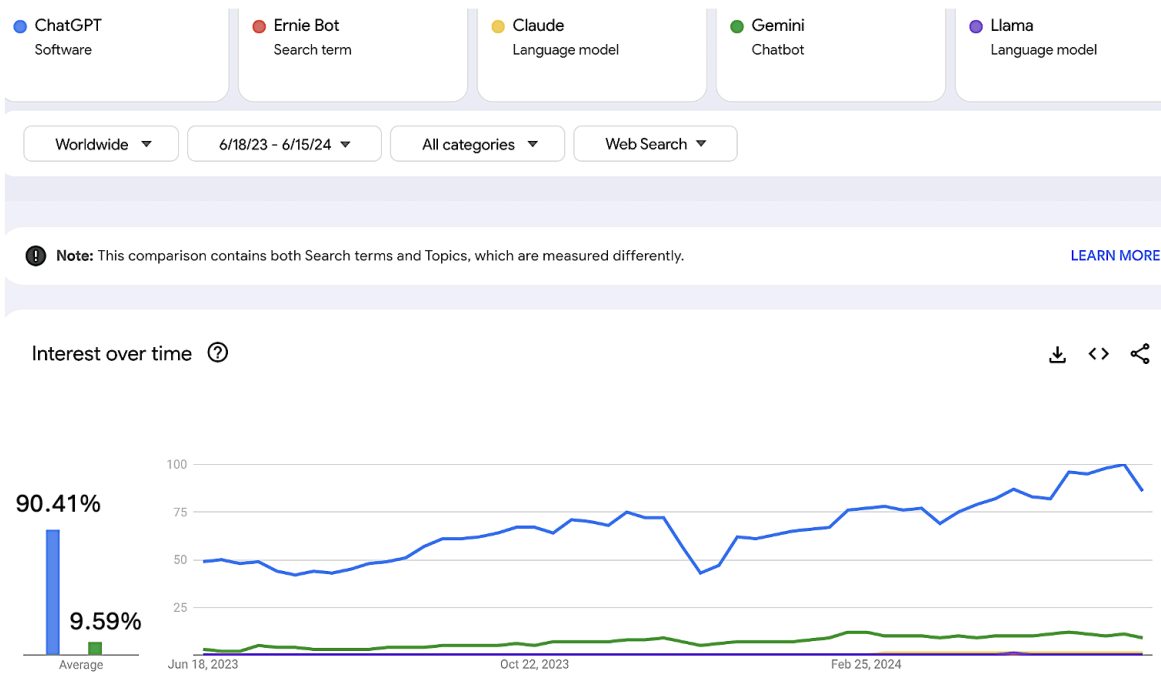

(b) ChatGPT dominates the attention of available LLMs, measured by Google Search.

**Figure S1.** *Search volumes in clean and full periods.*

## 2. Countries without legal access

OpenAI provides a list of supported countries for its ChatGPT and/or other services like the API, and this list has been periodically updated. We obtained historical versions of the list from the Internet Archive's Wayback Machine, [https://web.archive.org/web/20240000000000*/https://help.openai.com/en/articles/7947663-chatgpt-supported-countries](https://web.archive.org/web/20240000000000*/https://help.openai.com/en/articles/7947663-chatgpt-supported-countries). Using the first available list of supported countries, we identified all countries with more than ten preprints in our BioRxiv/MedRxiv and Arxiv datasets, both before and after the introduction of ChatGPT, and checked whether they were on the list. If they were not, we considered them initially prohibited, and tracked whether/when they gained or lost access using updates to the list and public news sources. The resulting list is displayed in Table S1 below. Preprints from the countries in S1 and countries with access account for 99.983% of the full BioRxiv corpus and 99.971% of the full Arxiv corpus. In the analysis of ChatGPT use and legal access we excluded preprints from countries that were prohibited at any point and had fewer than 10 preprints before and after ChatGPT.

English is not the predominant language in the countries listed in Table S1, as determined by their *de jure* official languages. In Cameroon, while English is co-official with French, it is spoken only in the Northwest and Southwest regions of the country[1].

---

[1] [https://translatorswithoutborders.org/language-data-for-cameroon](https://translatorswithoutborders.org/language-data-for-cameroon)

*Table S1*. *Major countries with restrictions on ChatGPT access in our data*

| Country/Region | Period of Restricted Access |
|---|---|
| Cameroon | 11/30/2022 - 11/06/2023[2] |
| China (Plus Hong Kong and Macau[3]) | 11/30/2022 - present |
| Egypt | 11/30/2022 - 11/01/2023[4] |
| Iran | 11/30/2022 - present |
| Russia | 11/30/2022 - present |
| Saudi Arabia | 11/30/2022 - 08/11/2023[5] |
| Vietnam | 11/30/2022 - 11/06/2023[6] |
| Ukraine (with certain exceptions) | 11/30/2022 - 2/18/2023[7] |
| Italy | 3/31/2023 - 4/28/2023[8] |

---

[2] https://web.archive.org/web/20231106222038/https://help.openai.com/en/articles/7947663-chatgpt-supported-countries . Accessed 2024-06-14.
[3] https://www.rfa.org/cantonese/firewall_features/firewall-chatgpt-04072023092323.html Accessed 2024-06-14.
[4] https://scenenow.com/Buzz/ChatGPT-is-Now-Officially-Available-in-Egypt. Accessed 2024-06-14.
[5] https://www.ngmisr.com/en/tech/7490. Accessed 2024-06-14.
[6] https://web.archive.org/web/20231106222038/https://help.openai.com/en/articles/7947663-chatgpt-supported-countries . Accessed 2024-06-14.
[7] https://mezha.media/en/2023/02/18/chatgpt-finally-started-working-in-ukraine/ Accessed 2024-06-14. The lifting of the ban did not apply to territories occupied by Russia.
[8] https://www.dw.com/en/ai-italy-lifts-ban-on-chatgpt-after-data-privacy-improvements/a-65469742 Accessed 2024-06-14.

# 3. Composition of no-access countries in Arxiv and BioRxiv corpora

In the Main paper, we disaggregated non-access countries into China and others, because China dominates the preprint volume in this group. Figures S2 and S3, for BioRxiv and Arxiv respectively, show that the top three countries in terms of productivity are China, Russia, and Iran, and together they add up to almost 100%.

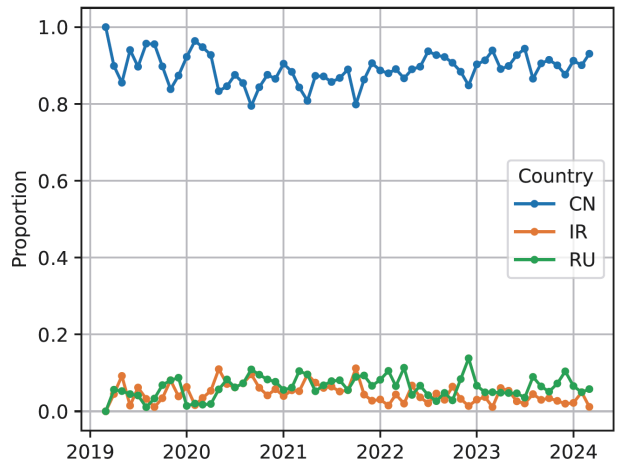

**Figure S2**: *BioRxiv preprints from no-access countries predominantly come from China, Russia, and Iran. Here, we define a preprint as coming from a certain country if all authors are from that country. Operationalizing it via the first or last author shows qualitatively identical patterns.*

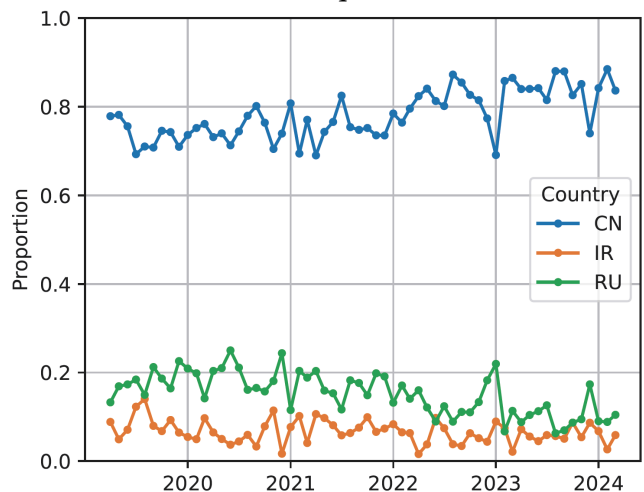

**Figure S3**: *Arxiv preprints from no-access countries predominantly come from China, Russia, and Iran. Here, we define a preprint as coming from a certain country if all authors are from that country. Operationalizing it via the first or last author shows qualitatively identical patterns.*

# 4. Descriptive analysis of Arxiv corpus

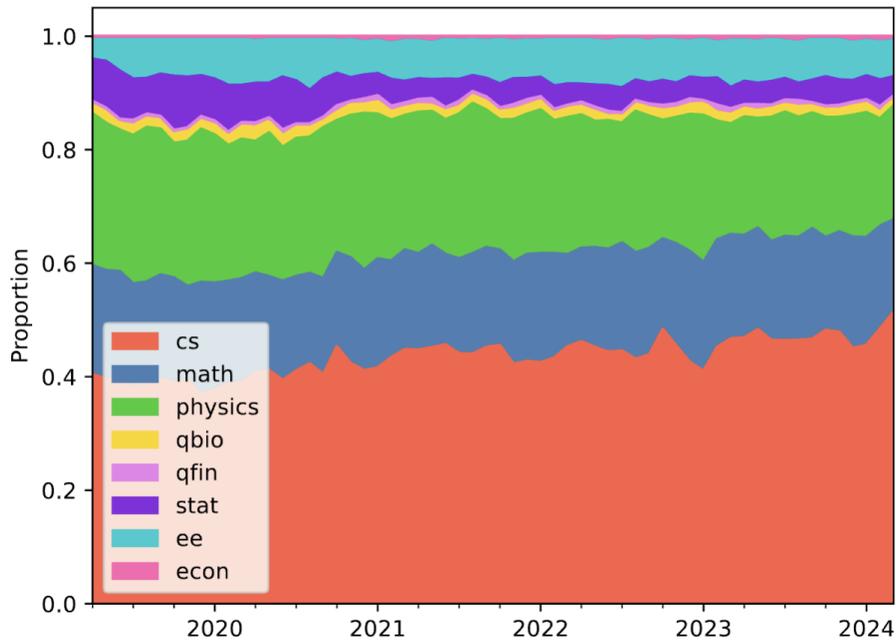

**Figure S4.** *Normalized composition of the Arxiv corpus by topic: computer science (cs) 43.43%, mathematics (math) 17.89%, physics 23.65%, quantitative biology (qbio) 1.61%, quantitative finance (qfin) 0.76%, statistics (stat) 5.05%, electronic engineering (ee) 6.99%, economics (econ) 0.63%. The raw labels of papers are not exclusive, meaning a single paper can have multiple subject labels. As a result, the sum of raw labels' proportions exceeds 1. In this figure we show the normalized proportion of each subject so that their sum equals one, and use the raw labels in the regression analysis.*

# 5. Attention, Citation, and Impact Factors

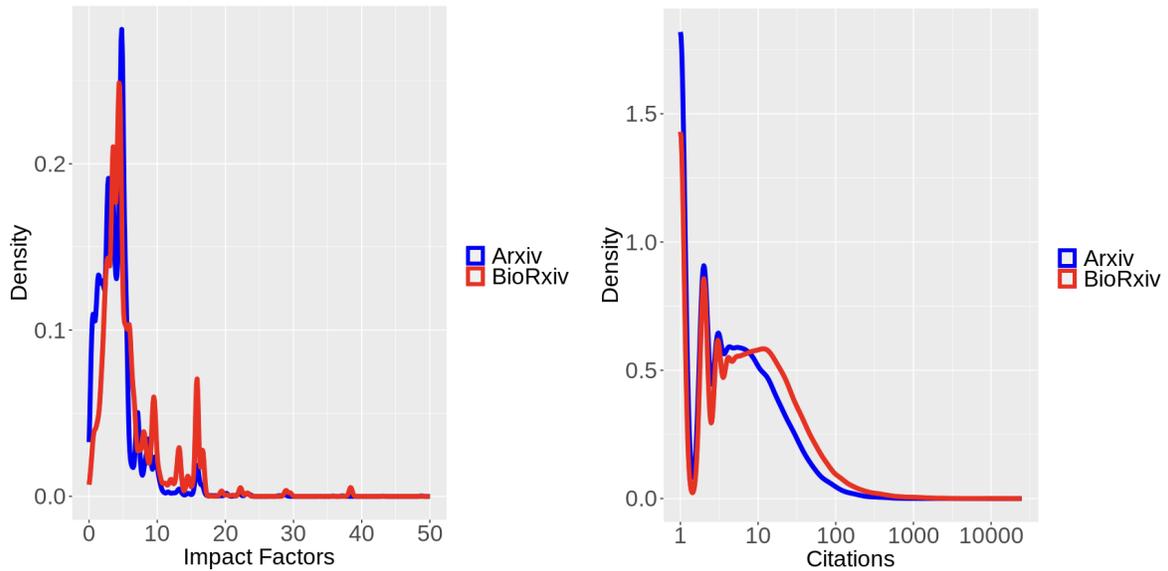

(a) Raw impact factors of ultimate publications and cumulative citations of preprints/published papers in Arxiv and BioRxiv.

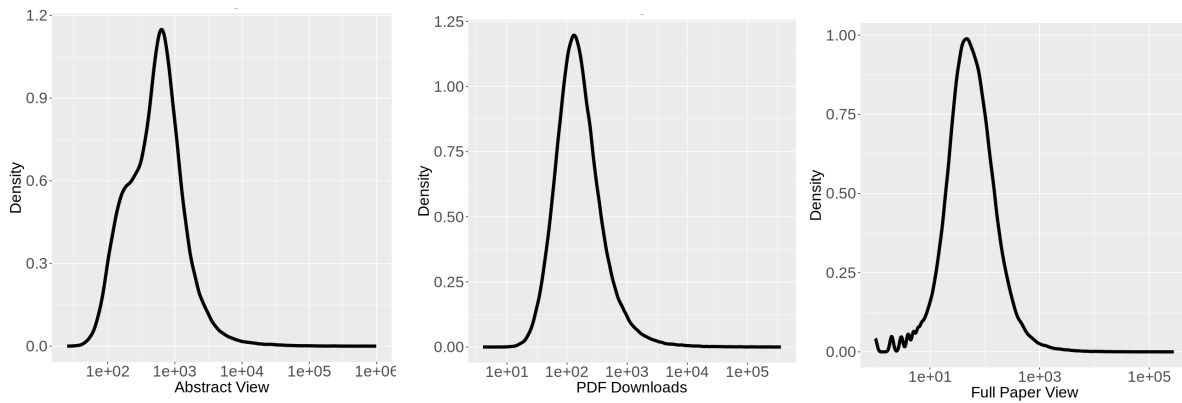

(b) Raw attention measures (6-month abstract views, PDF downloads, and full paper online views) of preprints in BioRxiv. Arxiv does not provide attention data.

**Figure S5**: *Skewed distributions of attention, citation, and impact factors.*

# 6. Prompts used in the classifier

- Please polish the following abstract. Be compelling: 
- Help me polish the following abstract with minimum changes: 
- Please polish the following abstract. Be clear: 
- We are writing a manuscript and plan to submit it to a top venue in our field. Below is our abstract. Please rewrite the abstract: 
- Please polish the following abstract. Be concise: 
- Rewrite the following abstract for a broad audience: 

In our experiments, prompts occasionally generate text enclosed in quotation marks (i.e., "<polished abstracts>"). Removing or retaining the quotation marks in the training set yields qualitatively similar classification results.

# 7. Classifier and off-the-shelf LLM-detectors performance

*Table S2.* *Performance metrics of our ensemble classifier versus GPTZero and ZeroGPT across 3 corpora. Each classifier was tested in 10,000 iterations of simulation on the three corpora from countries with/without access respectively. In each simulation, the proportion of label 1 in the sampled test set is randomized from 0-1. BioRxiv and Arxiv test sets are balanced with respect to texts from countries with and without legal access. Elsevier test set includes papers with self-declarations of AI use so at least one author in each paper has access to AI tools. The **bold** number is the best performance in a specific measure.*

| Testing sets | Measures | Ensemble | GPTZero | ZeroGPT |
|---|---|---|---|---|
| BioRxiv | $N$ | 20,000 | 20,000 | 20,000 |
| | Accuracy | **0.9710 ± 0.0001** | 0.9588 ± 0.0001 | 0.6060 ± 0.0002 |
| | F1 | **0.9447 ± 0.0006** | 0.9360 ± 0.0005 | 0.5424 ± 0.0014 |
| | Precision | 0.9375 ± 0.0008 | **0.9469 ± 0.0008** | 0.5731 ± 0.0020 |
| | TP (Recall) | **0.9626 ± 0.0001** | 0.9343 ± 0.0002 | 0.6048 ± 0.0005 |
| | TN | 0.9794 ± 0.0001 | **0.9835 ± 0.0001** | 0.6077 ± 0.0005 |
| | FP | 0.0206 ± 0.0001 | **0.0165 ± 0.0001** | 0.3923 ± 0.0005 |
| | FN | **0.0374 ± 0.0001** | 0.0657 ± 0.0002 | 0.3952 ± 0.0005 |
| | Gap | **0.0162 ± 0.0001** | 0.0291 ± 0.0002 | 0.2008 ± 0.0009 |
| | ROC-AUC | **0.9710 ± 0.0001** | 0.9589 ± 0.0001 | 0.6063 ± 0.0002 |
| Arxiv | $N$ | 20,000 | 20,000 | 20,000 |
| | Accuracy | **0.9442 ± 0.0001** | 0.8807 ± 0.0005 | 0.5453 ± 0.0018 |
| | F1 | **0.9046 ± 0.0008** | 0.8528 ± 0.0004 | 0.1691 ± 0.0003 |
| | Precision | 0.9009 ± 0.0011 | **0.9622 ± 0.0006** | 0.9124 ± 0.0012 |
| | TP (Recall) | **0.9275 ± 0.0002** | 0.7702 ± 0.0003 | 0.0946 ± 0.0002 |
| | TN | 0.9612 ± 0.0001 | 0.9916 ± 0.0001 | **0.9959 ± 0.0001** |
| | FP | 0.0388 ± 0.0001 | 0.0084 ± 0.0001 | **0.0041 ± 0.0001** |
| | FN | **0.0725 ± 0.0002** | 0.2298 ± 0.0003 | 0.9054 ± 0.0002 |
| | Gap | **0.0314 ± 0.0001** | 0.1112 ± 0.0005 | 0.4508 ± 0.0019 |
| | ROC-AUC | **0.9443 ± 0.0001** | 0.8809 ± 0.0001 | 0.5453 ± 0.0001 |
| Elsevier testing set | $N$ | 20,000 | 20,000 | 20,000 |
| | Accuracy | **0.9857 ± 0.0001** | 0.6182 ± 0.0022 | 0.5901 ± 0.0023 |
| | F1 | **0.9738 ± 0.0005** | 0.3758 ± 0.0014 | 0.3147 ± 0.0013 |
| | Precision | **0.9666 ± 0.0008** | 0.9516 ± 0.0022 | 0.8626 ± 0.0026 |
| | TP (Recall) | **0.9855 ± 0.0003** | 0.2408 ± 0.0011 | 0.2009 ± 0.0011 |
| | TN | 0.9854 ± 0.0003 | **1.0000 ± 0.0000** | 0.9863 ± 0.0003 |
| | FP | 0.0146 ± 0.0003 | **0.0000 ± 0.0000** | 0.0137 ± 0.0003 |
| | FN | **0.0145 ± 0.0003** | 0.7592 ± 0.0011 | 0.7991 ± 0.0011 |
| | Gap | **0.0111 ± 0.0001** | 0.3818 ± 0.0022 | 0.3962 ± 0.0024 |
| | ROC-AUC | **0.9855 ± 0.0002** | 0.6204 ± 0.0006 | 0.5936 ± 0.0006 |

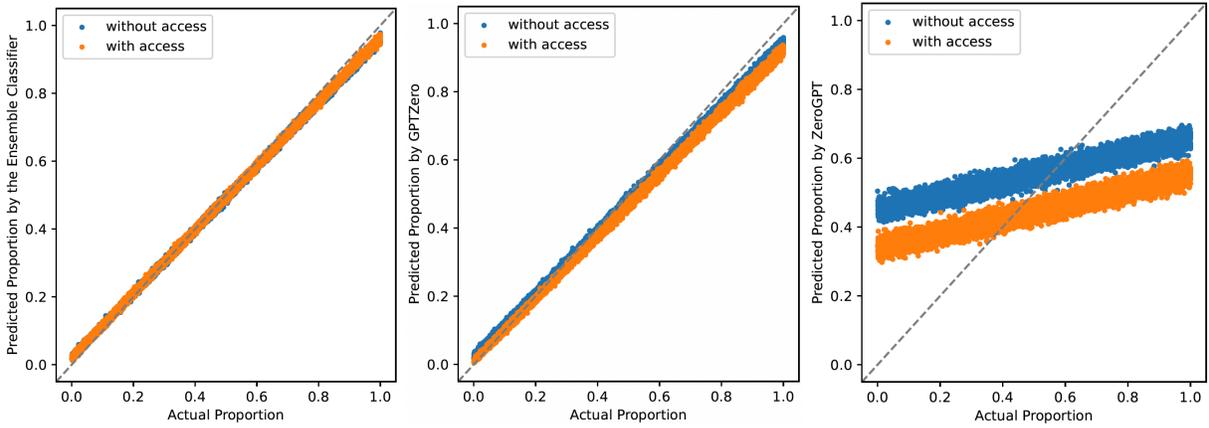

(a) On the BioRxiv Corpus

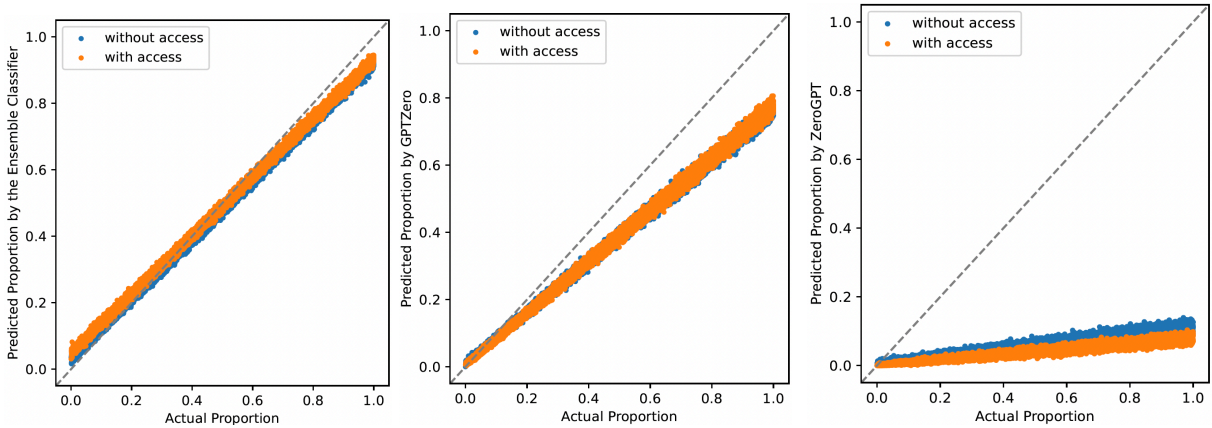

(b) On the Arxiv Corpus

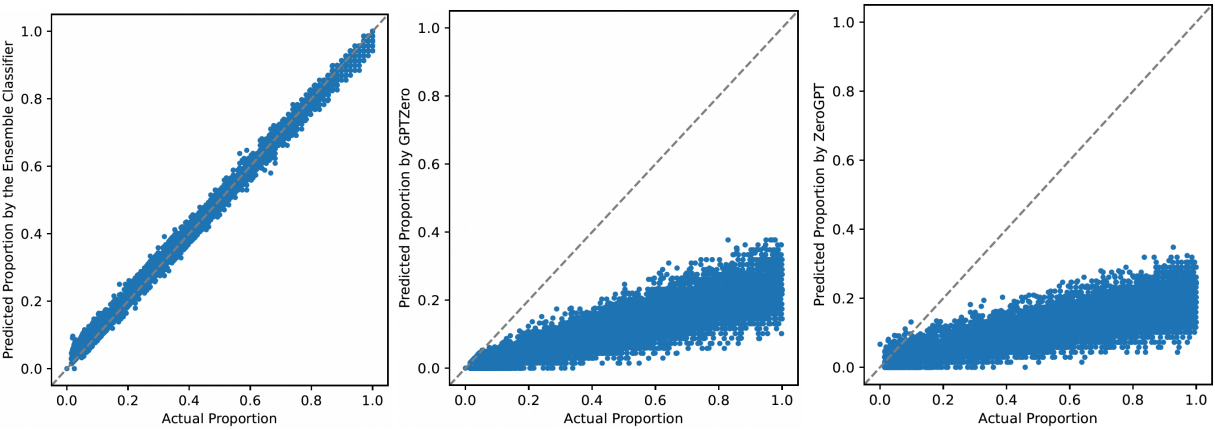

(c) On the Elsevier Corpus

*Figure S6. Performance Comparison of Classifiers.* The ensemble classifier demonstrates superior performance over both GPTZero and ZeroGPT in accurately estimating the proportion of ChatGPT-polished text across all three corpora from countries with and without legal access: BioRxiv (a), Arxiv (b), and Elsevier (c).

# 8. Mechanisms of the classifier using fingerprint words

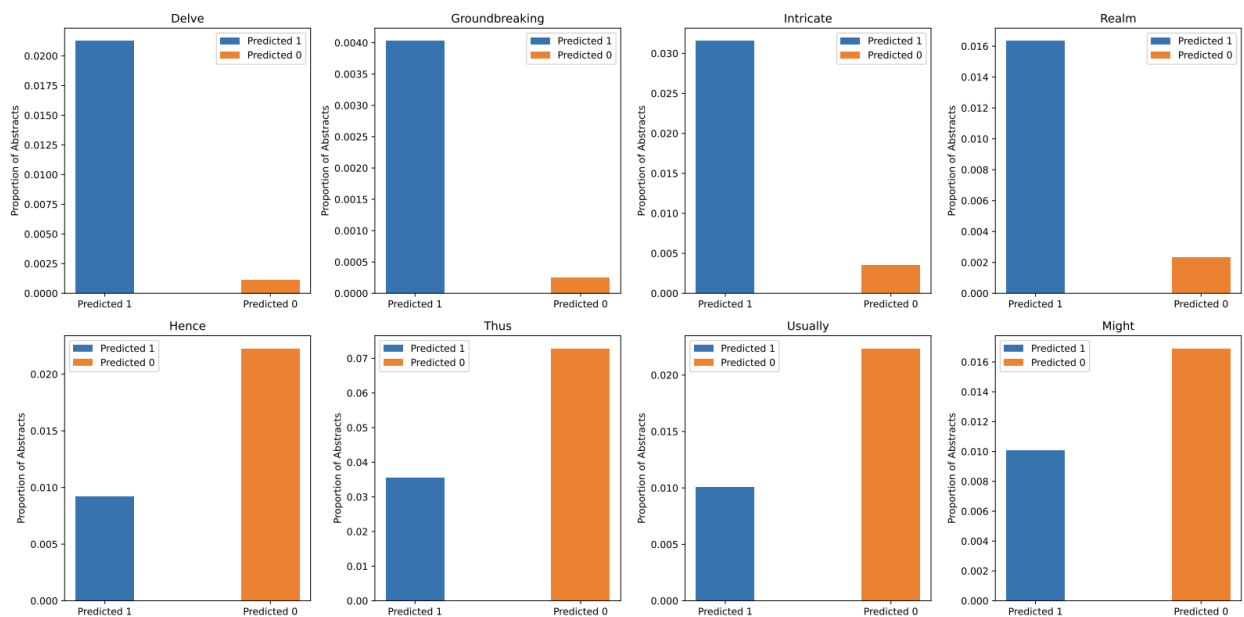

(a) Arxiv Corpus

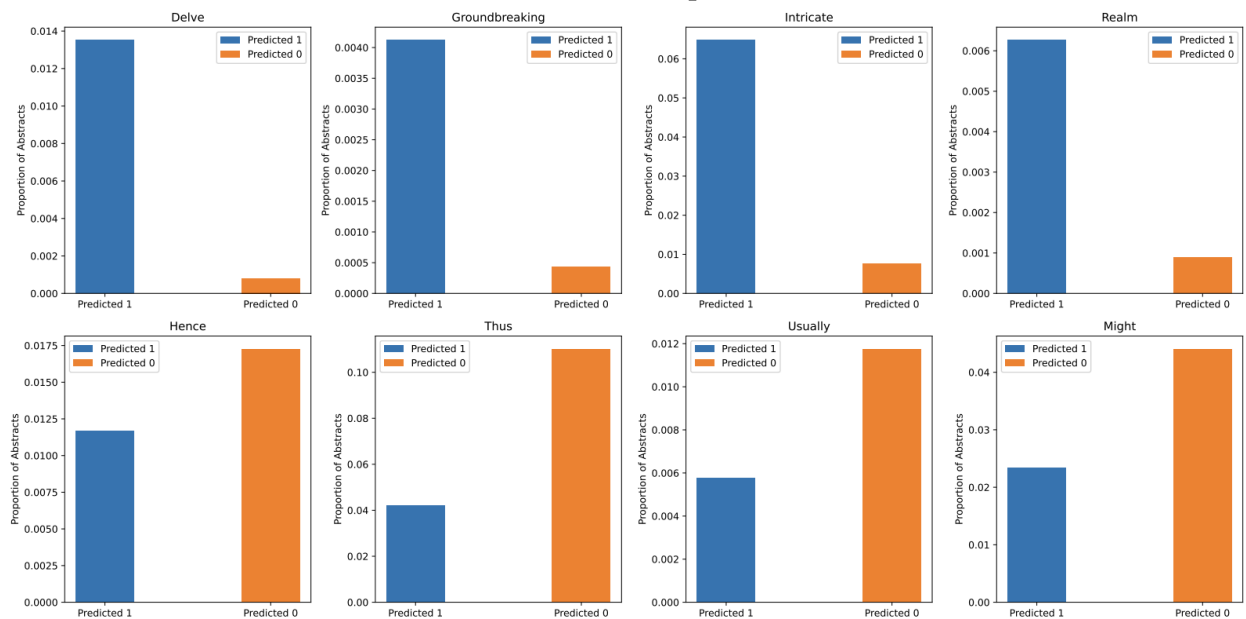

(b) BioRxiv Corpus

**Figure S7.** *Usage difference of 8 fingerprint words (Figure 1 of the Main text) in corpora labeled by the classifier as ChatGPT polished (blue bar) versus human-written (orange bar). All eight words show drastic differences in the two classes for both Arxiv corpora (top two rows) and BioRxiv corpora (bottom two rows). This suggests that the classifier may take advantage of fingerprint tokens for identifying the ChatGPT-polished text.*

# 9. Robustness check: alternative ways to define "legal access" for a preprint

In addition to the analysis in the Main text, which defines a paper as coming from a country without legal access if *all* authors are from such countries, we present two alternative definitions.

- First author-based: a paper is considered 'without legal access' if the first author is from a country without legal access, based on the assumption that the first author usually writes most of the paper. Our findings are consistent with the Main text results for both Arxiv and BioRxiv, as shown in Table S3.
- Last author-based: a paper is considered 'without legal access' if the last author is from a country without legal access, based on the assumption that the last author usually is the senior author who supervises the work, especially in domains like biological and computational sciences. Our findings are consistent with the Main text results for both Arxiv and BioRxiv, as shown in Table S4.

In both analyses standard errors are clustered at the last author level as in the Main text.

Table S3A. OLS models predicting use as a function of legal access of first author and other features in the BioRxiv corpus. Left column shows predictions for the full period and right column for the clean period.

|  | ChatGPT use (BioRxiv) | |
|---|---|---|
|  | Full | Clean |
| without access | 0.075*** | 0.006 |
|  | (0.005) | (0.007) |
| intercept | 0.012*** | 0.025*** |
|  | (0.003) | (0.003) |
| is computational | 0.059*** | 0.012* |
|  | (0.004) | (0.005) |
| year-month indicators | Y | Y |
| Observations | 65909 | 8432 |
| $R^2$ | 0.034 | 0.001 |
| Adjusted $R^2$ | 0.033 | 0.000 |
| Residual Std. Error | 0.284 (df=65890) | 0.166 (df=8427) |
| F Statistic | 112.951*** (df=18; 65890) | 1.587 (df=4; 8427) |

Note: *p<0.05; **p<0.01; ***p<0.001

Table S3B. OLS models predicting use as a function of legal access of first author and other features in the Arxiv corpus. Left column shows predictions for the full period and right column for the clean period.

|  | ChatGPT use (Arxiv) | |
|---|---|---|
|  | Full | Clean |
| without access | 0.028*** | -0.025* |
|  | (0.007) | (0.011) |
| intercept | 0.049*** | 0.062*** |
|  | (0.009) | (0.011) |
| computer science | 0.083*** | 0.035*** |
|  | (0.005) | (0.009) |
| economics | 0.005 | 0.064 |
|  | (0.037) | (0.089) |
| electronic engineering | 0.005 | 0.003 |
|  | (0.011) | (0.021) |
| mathematics | -0.083*** | -0.023 |
|  | (0.007) | (0.013) |
| physics | -0.073*** | -0.029** |
|  | (0.006) | (0.011) |
| quantitative biology | 0.028 | -0.051** |
|  | (0.019) | (0.017) |
| quantitative finance | 0.121** | 0.014 |
|  | (0.039) | (0.075) |
| statistics | 0.019 | -0.016 |
|  | (0.013) | (0.021) |
| year-month indicators | Y | Y |
| Observations | 18125 | 2576 |
| $R^2$ | 0.068 | 0.014 |
| Adjusted $R^2$ | 0.067 | 0.010 |
| Residual Std. Error | 0.353 (df=18099) | 0.239 (df=2564) |
| F Statistic | 51.128*** (df=25; 18099) | 3.205*** (df=11; 2564) |

Note: *p<0.05; **p<0.01; ***p<0.001

**Table S4A. OLS models predicting** use as a function of legal access of last author and other features in the BioRxiv corpus. Left column shows predictions for the full period and right column for the clean period.

|  | ChatGPT use (BioRxiv) | |
|---|---|---|
|  | Full | Clean |
| without access | 0.077*** | 0.010 |
|  | (0.006) | (0.008) |
| intercept | 0.012*** | 0.025*** |
|  | (0.003) | (0.003) |
| is computational | 0.059*** | 0.012* |
|  | (0.004) | (0.005) |
| year-month indicators | Y | Y |
| Observations | 65927 | 8432 |
| $R^2$ | 0.033 | 0.001 |
| Adjusted $R^2$ | 0.033 | 0.001 |
| Residual Std. Error | 0.284 (df=65908) | 0.166 (df=8427) |
| F Statistic | 112.997*** (df=18; 65908) | 1.819 (df=4; 8427) |

Note: *p<0.05; **p<0.01; ***p<0.001

**Table S4B. OLS models predicting** use as a function of legal access of last author and other features in the Arxiv corpus. Left column shows predictions for the full period and right column for the clean period.

|  | ChatGPT use (Arxiv) | |
|---|---|---|
|  | Full | Clean |
| without access | 0.028*** | -0.029** |
|  | (0.007) | (0.011) |
| intercept | 0.054*** | 0.063*** |
|  | (0.009) | (0.011) |
| computer science | 0.077*** | 0.028** |
|  | (0.006) | (0.010) |
| economics | 0.030 | 0.042 |
|  | (0.039) | (0.081) |
| electronic engineering | 0.005 | 0.005 |
|  | (0.012) | (0.022) |
| mathematics | -0.090*** | -0.024 |
|  | (0.007) | (0.013) |
| physics | -0.077*** | -0.027* |
|  | (0.006) | (0.011) |
| quantitative biology | 0.039 | -0.047* |
|  | (0.021) | (0.019) |
| quantitative finance | 0.106** | 0.037 |
|  | (0.039) | (0.081) |
| statistics | 0.009 | -0.000 |
|  | (0.013) | (0.023) |
| year-month indicators | Y | Y |
| Observations | 17016 | 2415 |
| $R^2$ | 0.069 | 0.013 |
| Adjusted $R^2$ | 0.068 | 0.008 |
| Residual Std. Error | 0.352 (df=16990) | 0.237 (df=2403) |
| F Statistic | 49.011*** (df=25; 16990) | 2.877*** (df=11; 2403) |

Note: *p<0.05; **p<0.01; ***p<0.001

# 10. Robustness check: alternative ways to specify panel regression for predicting academic outcomes

We consider an alternative specification of author fixed effects by focusing on the first author and clustering standard errors within the first author instead of the last author. As shown in the tables below, the results are broadly similar to those in the Main text, which uses last-author fixed effects. This analysis is limited to first authors with publications before and after the inception of ChatGPT, resulting in smaller sample sizes since junior authors tend to have lower productivity compared to senior authors. Consequently, some coefficients become insignificant.

**Table S5A. Panel OLS regressions predicting** attention (full paper online view, PDF downloads, and abstract view) as a function of ChatGPT use, time, and first author fixed effects. The first three columns use the full period and the last three the clean period. Attention measures are normalized by converting to percentile ranking within papers from the same field and time period.

|  | Full View Full Period | PDF Full Period | Abstract Full Period | Full View Clean Period | PDF Clean Period | Abstract Clean Period |
|---|---|---|---|---|---|---|
| used ChatGPT× post-ChatGPT | 0.026* (0.013) | 0.013 (0.012) | 0.015 (0.013) | 0.088* (0.038) | 0.037 (0.042) | 0.037 (0.043) |
| post-ChatGPT | -0.005 (0.004) | -0.008* (0.003) | -0.004 (0.004) | -0.043*** (0.007) | -0.065*** (0.007) | -0.069*** (0.008) |
| First author-FE | Y | Y | Y | Y | Y | Y |
| Observations | 25472 | 25693 | 25693 | 6373 | 6436 | 6436 |
| N. of groups | 9194 | 9234 | 9234 | 2259 | 2268 | 2268 |
| $R^2$ | 0.000 | 0.000 | 0.000 | 0.009 | 0.020 | 0.020 |
| Residual Std. Error | 0.005 (df=16276) | 0.005 (df=16457) | 0.003 (df=16457) | 0.025 (df=4112) | 0.037 (df=4166) | 0.039 (df=4166) |
| F Statistic | 2.461 (df=9196; 16276) | 3.029* (df=9236; 16457) | 1.020 (df=9236; 16457) | 18.566*** (df=2261; 4112) | 42.459*** (df=2270; 4166) | 42.633*** (df=2270; 4166) |

Note: *p<0.05; **p<0.01; ***p<0.001

**Table S5B. Panel OLS regressions predicting** citations and journal placement (if published in a journal) as a function of ChatGPT use, time, and first author fixed effects. The first four columns use the full period and the last four the clean period. Both citations and impact factors (IF) are normalized by converting to percentile ranking within papers from the same field and time period.

|  | IF Full Period BioRxiv | Citation Full Period BioRxiv | IF Full Period Arxiv | Citation Full Period Arxiv | IF Clean Period BioRxiv | Citation Clean Period BioRxiv | IF Clean Period Arxiv | Citation Clean Period Arxiv |
|---|---|---|---|---|---|---|---|---|
| used ChatGPT× post-ChatGPT | -0.025 (0.031) | -0.007 (0.008) | 0.024 (0.061) | -0.026** (0.010) | 0.128 (0.091) | 0.056 (0.035) | 0.053 (0.027) | -0.032 (0.057) |
| post-ChatGPT | 0.003 (0.007) | -0.034*** (0.003) | -0.036** (0.013) | -0.008 (0.004) | -0.002 (0.012) | -0.028*** (0.007) | -0.026 (0.024) | -0.017 (0.011) |
| First author-FE | Y | Y | Y | Y | Y | Y | Y | Y |
| Observations | 6311 | 37833 | 1289 | 12177 | 1852 | 6443 | 368 | 2108 |
| N. of groups | 2385 | 13533 | 523 | 4379 | 713 | 2270 | 161 | 765 |
| $R^2$ | 0.000 | 0.007 | 0.009 | 0.002 | 0.002 | 0.004 | 0.006 | 0.002 |
| Residual Std. Error | 0.004 | 0.020 | 0.022 | 0.009 | 0.010 | 0.016 | 0.017 | 0.011 |
| F Statistic | 0.395 | 82.712*** | 3.569* | 6.917*** | 0.904 | 8.617*** | 0.638 | 1.432 |
| df | 3924 | 24298 | 764 | 7796 | 1137 | 4171 | 205 | 1341 |

Note: *p<0.05; **p<0.01; ***p<0.001

# 11. ChatGPT use in Asian countries with similar expected demand

Figure S8 and Table S6 illustrate the usage patterns in Asian countries (excluding Singapore, Philippines, India, and Pakistan where English is an official or educational language). We specifically focus on these Asian countries due to their substantial linguistic differences from English, suggesting a higher potential benefit from ChatGPT's writing capabilities and higher expected demand. We have merged the corpora from Arxiv and BioRxiv due to the notably smaller volume of preprints. Consistent with findings in the Main text, countries lacking legal access demonstrate higher usage throughout the full period.

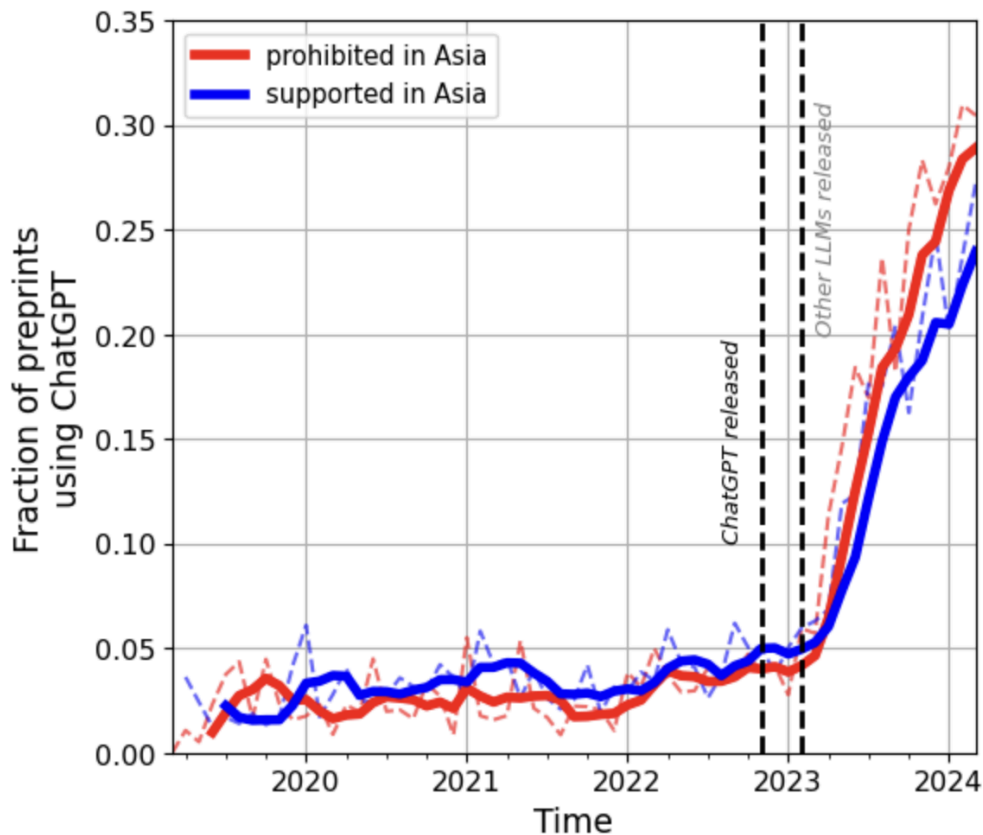

**Figure S8.** *Usage patterns in Asian countries where English is not an official language.*

*Table S6. OLS models predicting* use as a function of legal access and other features in the combined corpus of BioRxiv and Arxiv. The variable "without_access" indicates that no author had access to ChatGPT. Standard errors are clustered at the last author level. Left column shows predictions for the full period and right column for the clean period.

|  | ChatGPT use (BioRxiv + Arxiv) | |
|---|---|---|
|  | Full | Clean |
| without access | 0.022** | -0.007 |
|  | (0.008) | (0.011) |
| intercept | 0.003 | 0.031** |
|  | (0.009) | (0.010) |
| computer science | 0.108*** | 0.041* |
|  | (0.011) | (0.020) |
| economics | 0.167 | 0.026 |
|  | (0.142) | (0.017) |
| electronic engineering | -0.026 | -0.003 |
|  | (0.024) | (0.047) |
| mathematics | -0.075*** | 0.001 |
|  | (0.013) | (0.023) |
| physics | -0.039*** | -0.002 |
|  | (0.012) | (0.019) |
| quantitative biology | 0.090 | 0.115 |
|  | (0.063) | (0.160) |
| quantitative finance | 0.129 | -0.024 |
|  | (0.102) | (0.021) |
| statistics | 0.058 | -0.052*** |
|  | (0.041) | (0.014) |
| computational BioRxiv | 0.082*** | -0.011 |
|  | (0.013) | (0.014) |
| year-month indicators | Y | Y |
| Observations | 9852 | 1206 |
| $R^2$ | 0.068 | 0.011 |
| Adjusted $R^2$ | 0.065 | 0.001 |
| Residual Std. Error | 0.358 (df=9825) | 0.194 (df=1193) |
| F Statistic | 30.105*** (df=26; 9825) | 4.101*** (df=12; 1193) |

Note: $^*p<0.05$; $^{**}p<0.01$; $^{***}p<0.001$